# On the Analysis of Reed Solomon Coding for Resilience to Transient/Permanent Faults in Highly Reliable Memories

L.Schiano, M.Ottavi, F.Lombardi
Department of Electrical and Computer Engineering
Northeastern University,
Boston, MA 02115, USA

S.Pontarelli, A.Salsano
Department of Electronic Engineering
University of Rome "Tor Vergata"
Rome, Italy

## Abstract

*Single Event Upsets (SEU) as well as permanent faults can significantly affect the correct on-line operation of digital systems, such as memories and microprocessors; a memory can be made resilient to permanent and transient faults by using modular redundancy and coding. In this paper, different memory systems are compared: these systems utilize simplex and duplex arrangements with a combination of Reed Solomon coding and scrubbing. The memory systems and their operations are analyzed by novel Markov chains to characterize performance for dynamic reconfiguration as well as error detection and correction under the occurrence of permanent and transient faults. For a specific Reed Solomon code, the duplex arrangement allows to efficiently cope with the occurrence of permanent faults, while the use of scrubbing allows to cope with transient faults.*

*Index Terms: High Reliability Systems, Reliability Evaluation, Reed-Solomon Codes, Scrubbing, Dynamic Redundancy.*

## 1. Introduction

A generic approach to fault tolerance is seldom very efficient due to the diversity of applications and requirements as well as innovations in technology. For example, magnetic tape recorders have been extensively used to store the large amount of data commonly generated by the on-board instrumentation of satellites. However, their mechanical and electro-mechanical parts have insufficient flexibility to attain fault tolerance and reliability for long-term space missions. Moreover, the rapid growth in capacity of semiconductor memories and their compact size have permitted the development of Solid State Mass Memories (SSMMs), which are preferred over tape recorders due to their higher reliability [6]. However, the requirements of low latency, high throughput and large storage capacity of a SSMM may not be fully met by space certified components. Commercial Off The Shelf (COTS) offer better performance in terms of storage capacity, access speed and power consumption as compared with space-certified components. COTS however lacks resilience to both transient and permanent faults which commonly occur in a space environment. Memories are usually made fault tolerant by using a combination of Error Detection And Correction (EDAC) codes [3], and system-level techniques (such as sparing and scrubbing [2][10]). An EDAC code improves both the reliability of the storage system and data integrity (i.e. the probability that data is correctly stored in memory over a specified period of time); in this paper a class of maximum distance separable EDAC codes (known as Reed-Solomon codes, RS) is used. RS codes have been widely employed for transmission and storage due to their flexibility in the choice of dataword and codeword lengths. Modular sparing has been shown to improve the reliability of a memory system by replacing faulty modules or units (mostly affected by permanent faults). Scrubbing [2] which basically consists of periodically reading the content of the memory and correcting possible errors, improves data integrity by reducing the catastrophic accumulation of multiple transient errors (such as SEUs). Hence, the reliability of a memory system using these techniques is closely related to the occurrence of permanent faults (e.g. stuck-at, coupling etc), while data integrity is mainly related to the occurrence of transient faults resulting in errors (e.g. SEU), which can modify the value of the stored data.

The goal of this paper is to propose different arrangements (simplex and duplex) for a memory system which utilizes RS codes as EDAC; the objectives are to achieve resilience to both transient and permanent faults and retaining data integrity. In particular, the duplication of the memory modules (duplex) allows to cope with the occurrence of permanent faults, while the RS coding and the periodic execution of scrubbing allow to reduce the impact of transient faults. The performance of these fault tolerant memory systems is evaluated in terms of Bit Error Rate ($BER$) by Markov modeling for on-line reconfiguration.

The paper is organized as follows: Section 2 outlines the basic principles of RS codes, while in Section 3 the description of a simplex and the proposed duplex systems are provided. Preliminaries are given in Section 4. Section 5 illustrates the modeling method for estimating data integrity in both the simplex and the proposed duplex systems. Finally, in Section 6 analysis and conclusions are provided.

## 2 Review of Reed Solomon Coding

Reed Solomon codes [3] are widely used for both data transmission and storage systems. A RS($n,k$) code [3] is defined by the integer values of the two parameters $n$ and $k$, where $n$ denotes the number of symbols of $m$ bits (with $n \leq 2^m - 1$) of a codeword and $k$ denotes the number of symbols of the related dataword. A RS($n,k$) code can correct up to $2er + re \leq n - k$ erroneous bits, where $er$ is the number of erasures and $re$ is the number of random errors. For data transmission, a random error occurs when a symbol of the received codeword differs from the transmitted symbol in an unknown location of the codeword. An erasure is said to occur when the channel-side information (available from the receiver) allows to locate the erroneous symbol in the codeword. For a memory system, the following assumptions are applicable:
1) Transient faults (e.g. SEU) can occur in an unknown location (bit) of a codeword, therefore they can be effectively considered as random errors.
2) Permanent faults (e.g. stuck-at 0/1) can be located using either self-checking circuits, or on-line testing; therefore, they can be effectively considered as erasures.

Locating a permanent fault is necessary to exploit the error correction capability of the RS codes. Until the permanent fault is located, the error correction algorithm assumes the erroneous behavior to be caused by a random error, thus degrading the overall error correction capability of the provided code. When the permanent fault is located, the capability of the RS code can be fully exploited. The location of permanent faults can be achieved using different methods as found in the technical literature. For example, a permanent fault in a memory chip can be detected by monitoring the quiescent supply current ($I_{ddq}$) [9].

Furthermore, a technique known as scrubbing can be applied [2] to deal with the simulataneous presence (accumulation) of random errors in a codeword. Memory scrubbing basically consists of periodically reading a codeword, correcting the possible erroneous symbols and rewriting the corrected codeword to the same memory location. The usage of memory scrubbing must be carefully tuned to the system requirements as it also introduces some drawbacks. These are an increase of hardware overhead due to the necessary control circuitry, a reduction in memory availability during the scrubbing operations and an increase in power consumption.

## 3 Simplex and Duplex Configurations

In this paper two configurations for a memory system based on Reed-Solomon coding, are considered:
1) A simplex system makes use of a Reed-Solomon code and related co-decoder. In this configuration, $n$ coded symbols are read by the decoder and $k$ data symbols are provided as output.

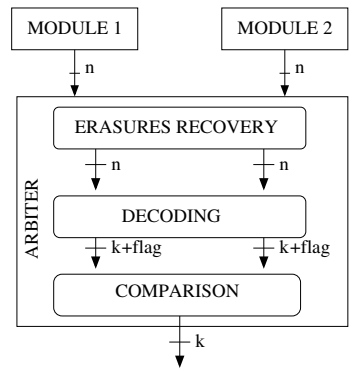

**Figure 1. Duplex system block diagram**

2) A duplex system with RS coding is shown in Fig. 1. This is based on the duplication of the memory module with an arbiter as decision circuit. This circuit operates as follows: initially, it attempts to recover from possible erasures occurring in either of the two words. Once the location of the erasures is established by the self-checking hardware, the arbiter masks erasures (provided that they do not occur in the same symbols of the words in the two modules). Following this operation, the arbiter separately decodes the two words to detect errors and correct them. Due to the limited correction capability of the code, mis-correction (i.e. correcting the erroneous word with yet another erroneous codeword) may occur. A flag is set when a correction has been performed and completed. Next, the arbiter proceeds by comparing the two words.

-If no flag is set, then one of the two words is provided as output (no error/fault present).

-If the two words are equal and at least one flag is set, then one of the two words is provided as output; in this case, the right correction has been performed. Note that the case of simultaneous mis-corrections (i.e. erroneous corrections leading to the same incorrect codeword) is not considered (masking error) due to its unlikely occurrence.

-If the two words are different and one of the two flags is set, then the word corresponding to the reset flag is provided as output. In this case, a mis-correction has occurred in the word corresponding to the module with the set flag.

-If the two words are different and both flags are set, the arbiter does not provide an output. Due to the duplex operation, the arbiter is not capable to discriminate between a correction and a mis-correction.

## 4 Preliminaries

Initially, the following definitions are introduced.
**Definition:** $\lambda$ is the fault exposure (or rate) of a single flipped bit (this is also known as SEU rate on a single bit).
**Definition:** $\lambda_e$ is the erasure exposure of a symbol.
**Definition:** $T_{sc}$ denotes the scrubbing operation period.
**Definition:** $BER$ is the Bit Error Rate i.e. the number of bits with errors divided by the total number of bits that have been read over a given time period (in percentage).

The systems considered in this paper have been modeled using Markov chains, which are particularly suitable for $BER$ evaluation [8]. To limit the problem of state explosion associated with Markov models of large and complex systems, only a word of a memory module (and its corresponding copy in the other module in the case of the duplex system) has been considered. However, the extension by considering the whole memory (memories) is straightforward and does not affect the ultimate correctness of the proposed models.

The following possible system operations and events have been considered as causing transitions in the Markov state diagrams:

1) A random error (bit-flip due to SEU) may occur in a word. This event leads the system to move to a neighboring state in which the effect of the bit-flip is considered.

2) Erasures have been considered over the symbols of the memory word encoded by the Reed Solomon code. Erasures cause the system to move to a neighboring state.

3) A scrubbing operation leads the system to a state characterized by the same number of permanent faults and no transient fault.

The following additional assumptions are made:
-When considering reliability, the inability to produce a correct output is considered to be a failure.
-Random errors on the same symbol are not considered.
-As in previous works in the technical literature, the arbiter (as decision circuit) is always assumed to be error-free (i.e. it is a hard core component).

## 5 Proposed Markov Models

Consider initially the simplex arrangement for the memory system, which has been introduced in [7]. The states $S(er, re)$ of the Markov Chain (CTMC) can be uniquely identified by the values of $er$ and $re$ The initial state at $T = 0$ (or Good state) is denoted as $G = S(0, 0)$; in this state $er = re = 0$. Similarly, the unrecoverable error state (or Fail state) $F = S(er, re)$ is a state in which $(er + 2 \times re) > n - k$.

The Markovian model of a RS coded simplex memory system capable to correct up to $t$ random errors (where $t = \frac{n-k}{2}$) is shown in Fig.2. Further details can be found in [7].

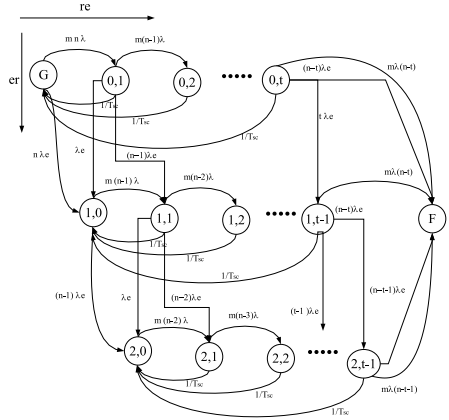

**Figure 2. Markov model of a RS coded simplex memory system**

According to the definition of $BER$ and by considering that $P_{S(n)}(t)$ is the time dependent solution of the Markov Chain (*i.e.*, the probability of being in state $S(n)$ at time $t$), the $BER$ of the system at time $t$ can be written as [7]:

$$BER(t) = m \times \frac{(n-k)}{k} \times P_{S(n)}(t) \quad (1)$$

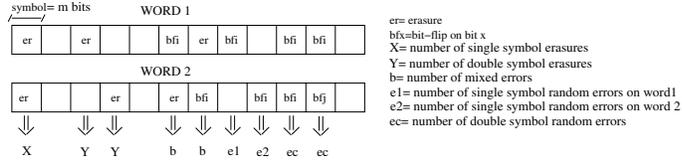

**Figure 3. State definition in the Markov chain**

Consider next the duplex memory system with the RS code. Six different parameters, depending on the erasures and random errors occurred, can be considered to characterize each state; they are defined in Fig.3, which shows two corresponding words in the replicated modules, as follows:

- *X* is the number of erasures on the same symbols in both words.
- *Y* is the number of erasures in symbols belonging to one of the two words, while the same symbol in the replicated module is error-free.
- *e1,e2* are the number of random errors (bit-flips) in either of the two words (word 1 or word 2 respectively), but

not in the same symbol of the other word.

- $ec$ is the number of random errors which have occurred in corresponding symbols of the two codewords (independently on the position of the bit-flip in the symbol itself).

- $b$ is the number of occurrences of an erasure and a random error in same symbols of the two replicated modules.

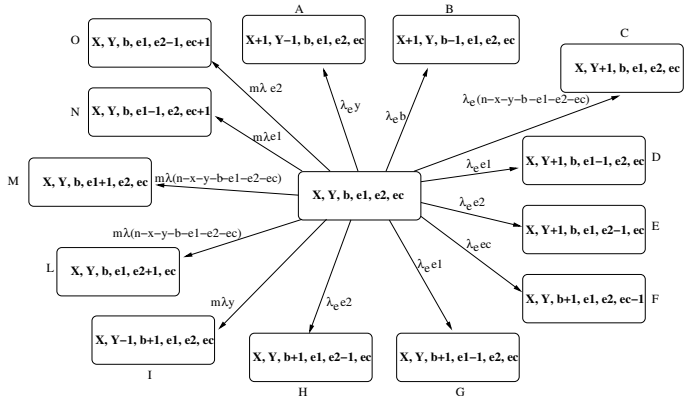

**Figure 4. Generic state of duplex system**

A state transition in the Markov chain model is caused both by events (i.e. random errors and erasures) and memory operations (scrubbing). Fig.4 shows all possible transitions from a generic state due to erasures and random errors. Note that the occurrence of a transition must satisfy the boundary conditions for the correct capabilities of the code employed. Therefore, from a generic state (characterized by the 6-tuple $(X, Y, b, e1, e2, ec)$) the possible transitions due to erasures are to the following states:

• $A$ is characterized by the 6-tuple $(X+1, Y-1, b, e1, e2, ec)$. Transitions to state $A$ are caused by an erasure on a symbol for which the same symbol in the duplicated module already has an erasure (at a rate $\lambda_e \cdot Y$).

• $B$ is characterized by the 6-tuple $(X+1, Y, b-1, e1, e2, ec)$. Transitions to state $B$ are caused by the occurrence of an erasure on a symbol already affected by a random error (bit-flip), and for which the same symbol in the duplicated module presents an erasure too. The transition rate is given by $\lambda_e \cdot Y$.

• $C$ is characterized by the 6-tuple $(X, Y+1, b, e1, e2, ec)$. Transitions to state $C$ are caused by the occurrence of an erasure on a symbol such that this symbol and the same symbol in the duplicated word were not previously affected by random error or erasures. The transition rate is $\lambda_e \cdot (n - X - Y - b - e1 - e2 - ec)$.

• $D, E$ are characterized by the 6-tuples $(X, Y+1, b, e1-1, e2, ec)$ and $(X, Y+1, b, e1, e2-1, ec)$, respectively. If an erasure occurs to a word previously corrupted by a random error (and whose homologous is error-free) then the system enters the $D$ or $E$ state. The transition rate is given by $\lambda_e \cdot e1$ or $\lambda_e \cdot e2$, respectively.

• $F$ is characterized by the 6-tuple $(X, Y, b+1, e1, e2, ec-1)$. A permanent fault occurring (with rate $\lambda_e \cdot ec$) to a symbol such that this symbol and the same symbol in the duplicated word were previously affected by random error takes the system to state $F$.

• $G, H$ are characterized by the 6-tuples $(X, Y, b+1, e1-1, e2, ec)$ and $(X, Y, b+1, e1, e2-1, ec)$, respectively. The system enters these states at the occurrence of an erasure on a previously error-free symbol, whose same symbol is affected by a random error. Transitions to $G$ or $H$ are characterized by a rate $\lambda_e \cdot e1$ or $\lambda_e \cdot e2$ respectively.

Similarly, if the boundary conditions are satisfied, the random errors (SEU causing bit-flips) cause the system to move to the following states:

• $I$ is characterized by the 6-tuple $(X, Y-1, b+1, e1, e2, ec)$. Transitions to state $I$ occur with rate $(m \cdot \lambda \cdot Y)$, where $m$ is the number of bits per symbol. They are caused by a random error occurring on a symbol, whose same symbol is affected by an erasure.

• $L, M$ are characterized by the 6-tuples $(X, Y, b, e1+1, e2, ec)$ and $(X, Y, b, e1, e2+1, ec)$, respectively. Transitions to these states occur following a random error on a symbol whose same symbol is error-free. Their rate is given by $\lambda \cdot m \cdot (n - x - y - b - e1 - e2 - ec)$.

• $N, O$ are characterized by the 6-tuples $(X, Y, b, e1-1, e2, ec+1)$ and $(X, Y, b, e1, e2-1, ec+1)$, respectively. The occurrence of a random error on a symbol whose same symbol is already affected by a random error causes a transition to one of these states. The rates of these transitions are $m \cdot \lambda \cdot e1$ and $m \cdot \lambda \cdot e2$ respectively.

As per the analysis of $BER$ for data storage, two system operations have been considered, namely read and scrubbing. The ability of the system to provide a correct output as result of a read operation is limited on each module by the condition $(2re + er) \leq n - k$. In the proposed duplex system, prior to any reading/scrubbing an erasure recovery operation is performed; therefore, the total number of erasures $er$ is $X$ while erasures on only one of the two homologous symbols ($Y$) can thus be masked by the arbiter. As for the total random errors on each word, $re$ is given by the sum of $b$, $ec$, and $e1$ or $e2$ respectively. Eventually all possible transitions shown in Fig. 4 may occur to characterize the next system configuration. However, either of the following conditions must be satisfied:

$$\begin{cases} X + 2b + 2ec + 2e1 & \leq \quad n - k \\ X + 2b + 2ec + 2e2 & \leq \quad n - k \end{cases}$$

Otherwise, the system reaches an unrecoverable error state (or Fail state). Read operations are not explicitly considered in the proposed model: a read operation corresponds to the so-called stopping time of a performed simulation. If the system is then in the Fail state, the read operation is unsuccessful. Else, a correct output is provided.

Similarly, a scrubbing operation is not successful when it is performed from a state which does not satisfy the boundary conditions, leading the system to the Fail state. Otherwise, a scrubbing operation causes a transition to a state characterized by the same number of permanent faults and no random error (thus to a state characterized by the 6-tuple $(X, Y+b, 0, 0, 0, 0)$). Scrubbing is executed at a prescribed frequency characterized by a rate $\frac{1}{T_{sc}}$.

## 6 Analysis and Discussion

The solution and evaluation in terms of $BER(t)$ of the proposed Markov models have been accomplished by using the SURE solver[4]. In the evaluation, rates for transient fault were varied from $7.3 \times 10^{-7}$ errors/bit/day to a maximum of $1.7 \times 10^{-5}$. As the rate of permanent faults depends on the reliability of the memory chips, then it can be established using for example the models of [6], [1].

In particular, the following three cases are considered:
1) comparison of simplex and duplex RS(18,16) with no scrubbing and variable SEU rate;
2) analysis of duplex RS(18,16) for different $T_{sc}$ periods and under the worst case scenario of the SEU rate;
3) comparison of simplex and duplex RS(18,16), and simplex RS(18,16) under different rates for permanent faults.

For cases 1) and 2), it is assumed that data is stored in the memory for 2 days ($T_{st} = 48h$); therefore $BER$ has been calculated during this time interval. The first evaluation has been performed on the simplex and duplex systems with RS(18,16) without scrubbing, no permanent fault and at a rate for transient faults given by $\lambda \in [7.3 \times 10^{-7}, 1.7 \times 10^{-5}]$. The $BER$s of the two systems are reported in Figure 5 and Figure 6; the values for the $BER$ are in the same range for all considered transient fault rates (of course, the duplex arrangement is expected to be particularly resilient to permanent faults, as also shown later in this section).

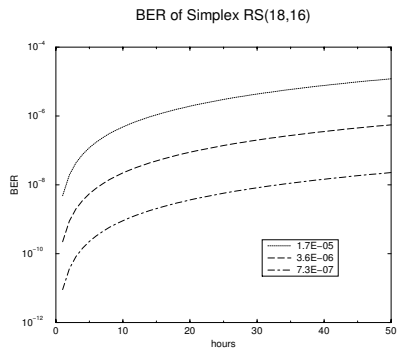

**Figure 5.** $BER$ **of simplex RS(18,16) under different SEU rates**

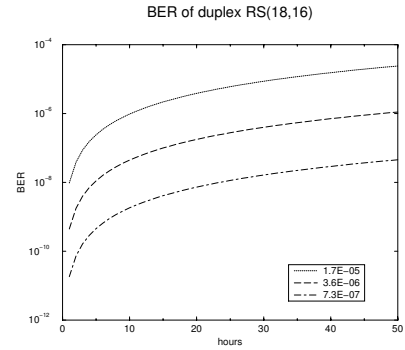

**Figure 6.** $BER$ **of duplex RS(18,16) under different SEU rates**

As for the duplex configuration, Figure 7 shows the improvement in $BER$ due to scrubbing. For this example a worst case transient faults scenario has been considered by setting $\lambda = 1.7 \times 10^{-5}$. Using a duplex system with the RS(18,16) code, a scrubbing frequency of lower than once per hour is sufficient to maintain the $BER$ below $10^{-6}$. This confirms previous results [7] on the simplex memory system using different RS codes.

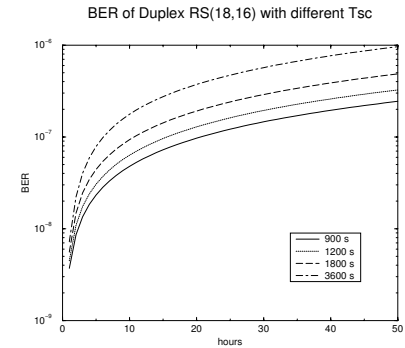

**Figure 7.** $BER$ **of duplex RS(18,16), fixed transient fault rate, variable scrubbing period**

Note that a RS(18,16) code in a duplex memory system implies the same amount of redundant code symbols as in a simplex system with a RS(36,16) code. Therefore, also this case is considered. Figures 8, 9, and 10 show the comparison of the RS(18,16) simplex, duplex and RS(36,16) simplex. These three cases are considered for the occurrence of permanent faults (over a period of 24 months for permanent storage of data in the memory). The RS(18,16) duplex has significantly better performance than the simplex memory, but it shows a degradation in performance compared with a simplex system employing a RS(36,16) code.

Consider next the decoder complexity. The time for Reed Solomon decoding ($T_d$) depends on the codeword length and on the number of code symbols as $T_d \simeq 3n +$ $10(n-k)$; this complexity has been reported in [5] for the performance of a RS codec IP core for FPGA implementation. The value of $T_d$ is in clock cycles for performing the decoding operation on a non time-continuous access profile as applicable to memory. Therefore, for a RS (36,16) decoder $T_d \simeq 108 + 200 = 308$ while for a RS (18,16) $T_d \simeq 54 + 20 = 74$, i.e. the decoding access time to the stored data is more than four times higher using the RS (36,16) arrangement than the simplex or duplex RS (18,16).

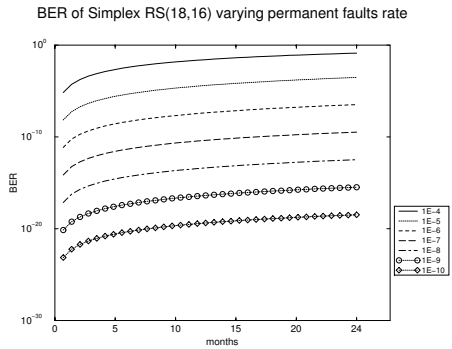

**Figure 8.** $BER$ **of simplex RS(18,16), different permanent fault rates, with no scrubbing.**

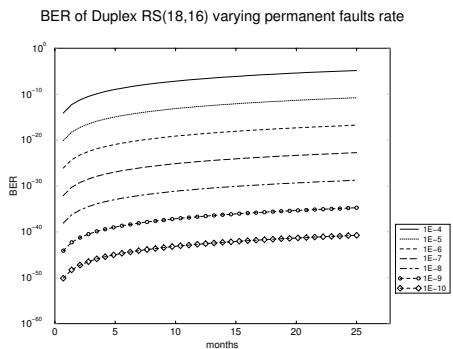

**Figure 9.** $BER$ **of duplex RS(18,16), different permanent fault rates, with no scrubbing.**

A further metric that has been evaluated, is the area of the decoder. The number of logic gates required for implementing a Reed Solomon decoder, is almost linearly dependent on $m$ and the number of check symbols $n - k$ [5]; therefore, a single RS(36,16) decoder will require more area than two RS(18,16) decoders for the duplex arrangement. The increase in area must be carefully evaluated when the physical constraints on electronic equipment are very tight; moreover, a substantial increase in decoder area also implies an increase in the non-fault tolerant components, thus ultimately degrading the overall reliability of the system.

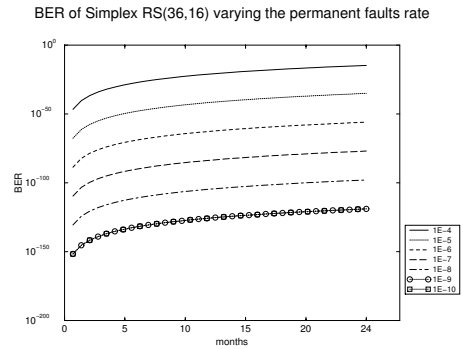

**Figure 10.** $BER$ **of simplex RS(36,16), different permanent fault rates, with no scrubbing.**

In conclusion, the novel Markov models presented in this paper have analytically shown that Reed Solomon codes can be efficiently employed in high reliable memory systems; albeit ultimately application dependent, redundancy and coding can be combined to attain data integrity and resilience to errors due to transient (SEU) and permanent faults. Different figures of merit such as $BER$ and decoding complexity (in time and area) have been analyzed in depth. The proposed models have provided an accurate and flexible evaluation tool which can be used to assess the viability of SSMMs for long mission time in space exploration.